\documentclass{article}
\usepackage{epsfig,psfig}
\usepackage{amsmath}
\usepackage{psfrag}
\usepackage{subfigure}
\usepackage{cite}

\begin{document}
\vspace{1.5cm}
\begin{center}
  {\Large \bf The pion form factor on the lattice at zero and finite 
temperature}\footnote{Presented by J. van der Heide at Light-Cone 2004, 
Amsterdam, 16 - 20 August}\\[.3cm] 
   \vspace{1.0cm}
   {\sc J. van der Heide$^a$, J.H. Koch$^a$, and E. Laermann$^b$}\\
   $^a${\it National Institute for Nuclear Physics and High-Energy Physics
   (NIKHEF)\\ 1009 DB Amsterdam, The Netherlands}\\
   $^b${\it Fakult\"at f\"ur Physik, Universit\"at Bielefeld,
     D-33615 Bielefeld, Germany}\\
 \end{center}
 \vspace{1cm}

\begin{abstract}
  We calculate the electromagnetic form factor of the pion in quenched
  lattice QCD. The non-perturbatively improved Sheikoleslami-Wohlert
  lattice action is used together with the consistently ${\mathcal
    O}(a)$ improved current. We calculate the pion form factor for
  masses down to $m_\pi = 360\; MeV$, extract the charge radius, and
  extrapolate toward the physical pion mass.  In the second part, we
  discuss results for the pion form factor and charge radius at $0.93
  \; T_c$ and compare with zero temperature results.
\end{abstract}
\section{Introduction} 
The pion, being the lightest and 'simplest' particle in the hadronic
spectrum has been studied intensively in the past. Using different
effective and phenemenological models, the properties of the pion have
been desribed with varying success.  However, these models make a
crucial assumption: confinement is put in by hand, in contrast to
being the result of the underlying dynamics.  Lattice QCD (LQCD) does
not have this drawback since it is solves QCD directly from first
principles.

Using LQCD, global properties of the pion such as the mass and the
decay width have been calculated to satisfying accuracy. The form
factor, which directly reflects the internal structure, is clearly an
important, additional challenge. The first lattice results were
obtained by Martinelli and Sachrajda \cite{Martinelli:1988bh}. It was
followed by a more detailed study by Draper {\it et al.}
\cite{Draper:1989bp}.  We extend these studies in several ways.
First, we adopt improved lattice techniques
\cite{Luscher:1997ug,Sheikholeslami:1985ij,Luscher:1997jn},
which means that we include extra operators in order to systematically
eliminate all the $\mathcal{O}(a)$ discretisation errors.
Furthermore, we extend the calculations to lower masses than achieved
before.

In view of heavy ion collisions, it is necessary to know whether, and
how, these non-perturbative observables change with temperature,
especially in the vicinity of the phase transition. To investigate
this, we have calculated the form factor at finite temperature for the
first time using lattice QCD. First results at $T=0.93 \; T_c$ are
presented in the second part.

\section{The method}

\begin{figure}
\subfigure[]{
  \includegraphics[width=0.45\textwidth]{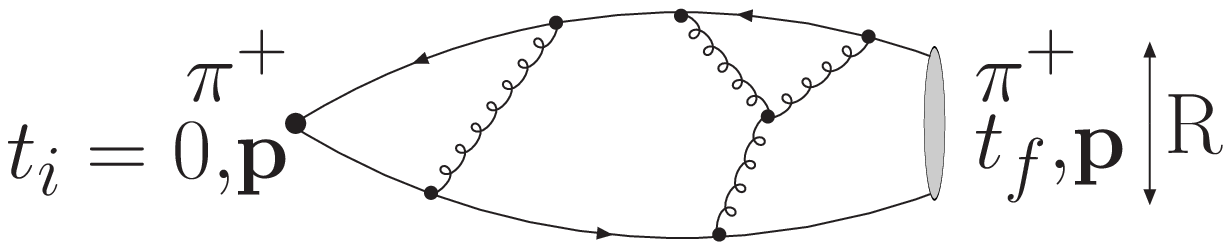}
  \label{fig:2p}
}
\subfigure[]{
  \includegraphics[width=0.45\textwidth]{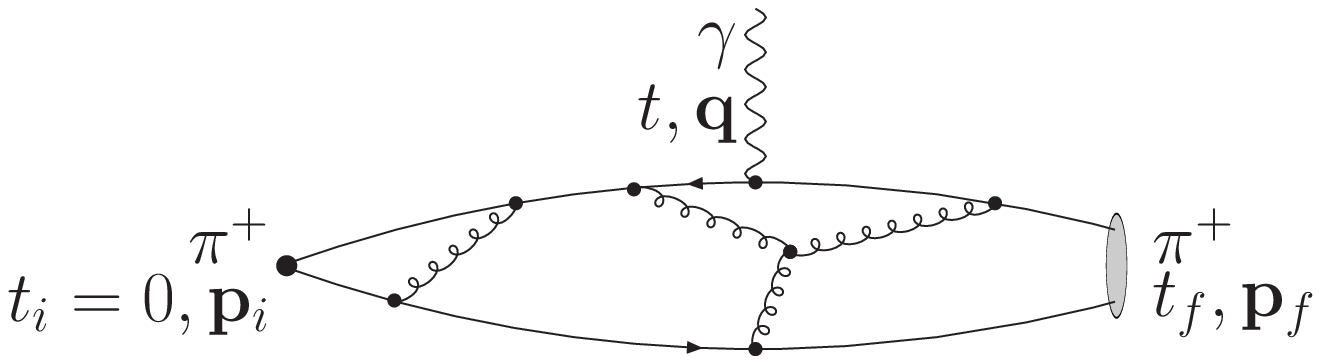}
  \label{fig:3p}
}
\caption{Graphic representation of (a) the two- and (b) the three-point function.}
\label{fig:2p_3p}
\end{figure}

The calculation of $n$-point functions is done using the Feynman path
integral approach. We exploit the similarity of the path integral with
the thermodynamic expectation value to define a temperature on the
lattice. The temperature is then defined as the inverse of the
temporal extent of the lattice. In order to reliably extract the form
factor, we need to evaluate both the two- and three-point Green's
function of the pion, shown in Fig. \ref{fig:2p_3p}.
For our form factor calculations we make use of non-perturbatively
improved, quenched Wilson fermions on a $24^3 \times 32$ lattice
($T=0$) and on a $32^3 \times 8$ lattice ($T=0.93\, T_c$). The gauge
coupling, $\beta=6.0$, corresponds to $g^2=1$ and a lattice spacing
$a=0.105$ fm. We generated $\mathcal{O}(100)-\mathcal{O}(200)$
configurations. Using these configurations, we calculated quark
propagators for five different quark masses. Details of our methods,
including analysis techniques, can be found in
\cite{vanderHeide:2003kh}. Here, we will only focus on the results and
the details of the finite temperature calculations.

In contrast to the $T=0$ case, for which we can write the pion-photon matrix element as 
\begin{equation}
\Gamma_{\mu}^{(T=0)}=\,(p_{\mathrm{f}}+p_{\mathrm{i}})_{\mu}\,F(Q^2)\, ,
\end{equation}
the parametrisation of the data is more difficult at finite
temperature. Here, the matrix element now involves three form
factors and they dependent on more scalar variables
\begin{multline}
\label{eq:ld_FT}
\Gamma_{\mu} = \,(p_{\mathrm{f}}+p_{\mathrm{i}})_{\mu}\,F(Q^2,{\bf p}^2_{\bot ,f},{\bf p}^2_{\bot ,\mathrm{i}},\omega_{n,\mathrm{f}},\omega_{n,\mathrm{i}}) 
+ \,q_{\mu}\,G(Q^2,{\bf p}^2_{\bot ,\mathrm{f}},{\bf p}^2_{\bot ,\mathrm{i}},\omega_{n,\mathrm{f}},\omega_{n,\mathrm{i}})\, \\
+ n_{\mu}\,H(Q^2,{\bf p}^2_{\bot ,\mathrm{f}},{\bf p}^2_{\bot ,\mathrm{i}},\omega_{n,\mathrm{f}},\omega_{n,\mathrm{i}})\, ,
\end{multline}
with ${\bf p}_{\bot}=(p_x, p_y)$ and $\omega_n$ are the Matsubara
frequencies. Using current conservation ($q_{\mu}\Gamma_{\mu}=0$), we
can remove one form factor. Imposing the extra kinematical restrictions
\begin{align}
{\bf p}_{\bot ,\mathrm{f}}^2 & ={\bf p}_{\bot ,\mathrm{i}}^2 \;\;\; \textrm{and}\\
\omega_{n, \mathrm{f}} &=\omega_{n, \mathrm{i}}=0 \, ,
\end{align}
and $\mu=3$, we are left with the form factor corresponding to $F$ in
the zero temperature case.  Furthermore, as can be seen, the form
factor may now depend on more scalar variables. The dependency on
${\bf p}_{\bot ,\mathrm{f}}^2$ and ${\bf p}_{\bot ,\mathrm{i}}^2$ will
be investigated here; $\omega_n$ will be chosen zero.

\section{Results}
As a byproduct of our simulations we also obtain pion masses for the 5
different $\kappa$-values. They agree with the literature. We also
checked the energy-momentum relation and up to the energies involved
we found that a continuum relation provides the best description.
These non-trivial tests indicate that our simulations are done
correctly and that we are close to the continuum limit.

\begin{figure}
\centering
    \psfrag{Q2 (GeV2)}[][]{\Huge $Q^2$ (GeV$^2$)}
    \psfrag{F(Q2)}[][]{\Huge $F(Q^2)$}
    \psfrag{mpi=970 MeV}{\Huge $m_{\pi}=970$ MeV}
    \psfrag{mpi=670 MeV}{\Huge $m_{\pi}=670$ MeV}
    \psfrag{mpi=360 MeV}{\Huge $m_{\pi}=360$ MeV}
    \psfrag{VMD, mV=690 MeV}{\Huge VMD, $m_V=690$ MeV}
    \psfrag{VMD, mV=721 MeV}{\Huge VMD, $m_V=721$ MeV}
    \psfrag{Experiment}{\Huge Experiment}
    \includegraphics[angle=-90,width=0.75\textwidth]{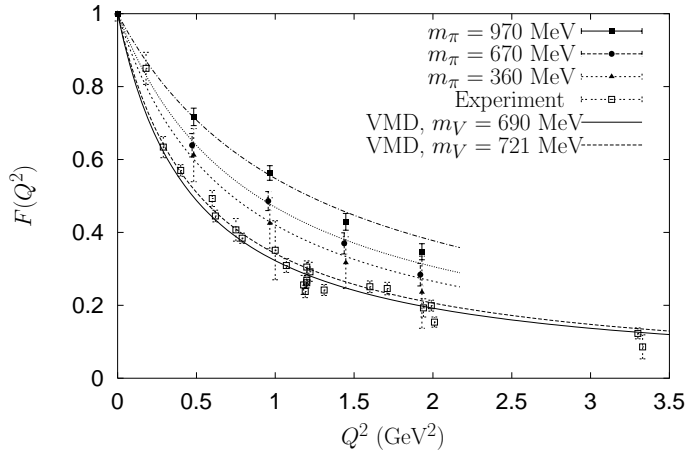}
  \caption{The form factor for different $m_{\pi}$, compared to experiment \protect\cite{Bebek:1978pe,Volmer:2000ek} ($T=0$)}
  \label{fig:FF_exp_comp}
\end{figure}

Using different currents, we extract the form factor for the five
$\kappa$-values at $T=0$. Comparison of the results for the Noether
current and the improved current shows that the effect of improvement
can be as large as 25$\%$ for the highest momentum transfers
considered.  The improved results are shown in
Fig.\ref{fig:FF_exp_comp}. The high accuracy of the data point at
$Q^2=0$ is due to the fact that the Ward-Takahashi identity related to
current conservation is satisfied to 1 ppm. From the figure, we
observe that our results approach the experimental data when the quark
mass is lowered.  Although our lightest pion is still more than twice
as heavy as the physical pion, we come rather close to the
measurements. As in the previous study \cite{Draper:1989bp} of the
pion form factor we have compared our results to a VMD inspired
monopole form. Fitting our data to this model, we extract a vector
meson mass which is within 5$\%$ of the corresponding rho mass on the
lattice \cite{Bowler:1999ae}.

\begin{figure}
\centering
  \psfrag{Q2}{\Huge $Q^2$}
  \psfrag{F(Q2)}{\Huge $F(Q^2)$}
  \psfrag{mpi=0.283}{\Huge $m_{\pi}=0.283$}
  \psfrag{P2=0.039}[][]{\Huge ${\bf p}^2_{\bot}=0.039$ }
  \psfrag{P2=0.077}[][]{\Huge ${\bf p}^2_{\bot}=0.077$ }
  \psfrag{P2=0.193}[][]{\Huge ${\bf p}^2_{\bot}=0.193$ }
  \includegraphics[angle=-90,width=0.75\textwidth]{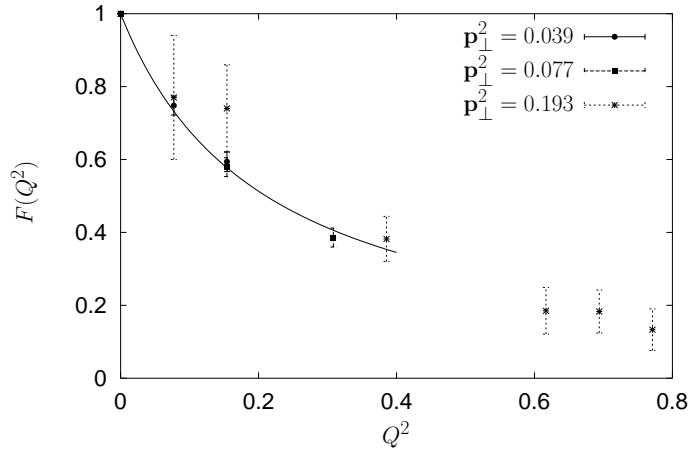}
  \caption{Form factor for different pion momenta, $m_{\pi}=530 $ MeV. 
    Curve: VMD fit to combined lower momenta.}
  \label{fig:FF_FT}
\end{figure}
The masses we extract from the two-point function at $T=0.93 \, T_c$
are slightly lower than at $T=0$. This effect is systematic, but the
error bars overlap. The data for finite momentum is best described by
the continuum dispersion relation. As for the $T=0$ data, the form
factor can be described by the VMD inspired monopole form with $m_{V}$
acting as a fit parameter up to moderate $Q^2$.  For one of our five
$\kappa$ values, the form factor is plotted for different pion momenta
in Fig.  \ref{fig:FF_FT}.  The form factor is seen to depend on the
pion momenta, although this dependence becomes important for higher
momenta only.  Since the lower momenta data seems to be
indistinguishable, we combine them to facilitate the VMD fit, this
combined set is denoted ${\bf p}_l$.
\begin{figure}[h]
\centering
  \psfrag{Q2}[][]{\Huge $Q^2$}
  \psfrag{F(Q2)}[][]{\Huge $F(Q^2)$}
  \psfrag{mpi=0.283}{\Huge $m_{\pi}=0.287$}
  \psfrag{T = 0.93 Tc, P2l}[][]{\Huge $T=0.93\; T_c$, ${\bf p}_{\mathrm{l}}$}
  \psfrag{T = 0.93 Tc, P2h}[][]{\Huge $T=0.93\; T_c$, ${\bf p}_{\mathrm{h}}$}
  \psfrag{T = 0}[][]{\Huge $T=0$}
  \psfrag{mpi=0.353}[][]{\Huge $m_{\pi}=660$ MeV}
  \includegraphics[angle=-90,width=0.75\textwidth]{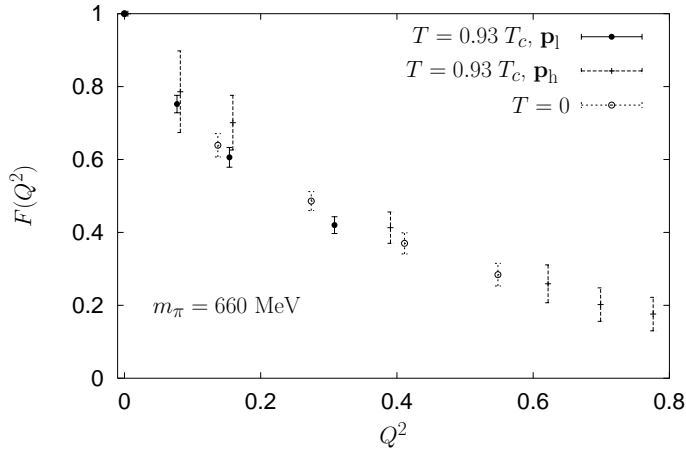}
\caption{Form factor as a function of $Q^2$ 
    for different T and different pion momenta.}
  \label{fig:FF_K13430_diff_T_comp}
\end{figure}

In Fig. \ref{fig:FF_K13430_diff_T_comp}, we compare the form
factor for different temperatures. It is seen that at $T=0.93 \, T_c$,
the lower momenta data fall off slightly faster than the $T=0$ data.
For ${\bf p}^2_{\bot}=0.193\, (={\bf p}_{\mathrm{h}})$, the opposite
occurs.

\begin{figure}
  \centering
  \psfrag{mpi2 (GeV2)}[][]{\Huge $m^2_{\pi}$ (GeV$^2$)}
  \psfrag{<r2> (fm2)}[][]{\Huge $\langle r^2 \rangle$ (fm$^2$)}
  \psfrag{0.93 Tc, Pl}[][]{\Huge $0.93 \; T_c$, ${\bf p}_{\mathrm{l}}$}
  \psfrag{0.93 Tc, Ph}[][]{\Huge $0.93 \; T_c$, ${\bf p}_{\mathrm{h}}$}
  \psfrag{T=0}[][]{\Huge $T\!\!=\!0$}
  \psfrag{VMD}[][]{\Huge VMD}
  \psfrag{Exp. value at T=0}[][]{\Huge Exp. value ($T\!\!=\!0$)}
  \includegraphics[angle=-90,width=0.75\textwidth]{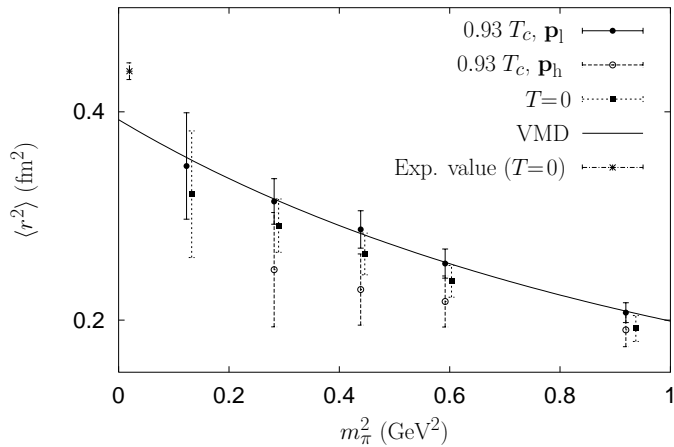}
  \caption{$\langle r^2 \rangle$ as a function of $m_{\pi}$ for different T, 
    and pion momenta. Curve: extrapolation based on VMD of the ${\bf p}_{\mathrm{l}}$ finite temperature results.}
  \label{fig:RMS}
\end{figure}

From the behaviour of the form factors at low $Q^2$, we can extract the
mean-square charge radius of the pion, which is shown in Fig.
\ref{fig:RMS} for both temperatures, together with an extrapolation of
the finite temperature results. The extrapolation method is not
unique; we have chosen to use the extrapolation based on VMD (see
\cite{vanderHeide:2003kh}) to obtain the radii in the physical limit.
The radii are
\begin{align}
\langle r^2 \rangle&=0.36(2)\;\; \textrm{fm}^2\textrm{ at}\;\;T=0\, ,\nonumber \\
\langle r^2 \rangle&=0.39(2)\;\; \textrm{fm}^2\textrm{ at}\;\;T=0.93\, T_c \;\; ({\bf p}_{\mathrm{l}})\;\; \textrm{and}\nonumber \\
\langle r^2 \rangle&=0.28(3)\;\; \textrm{fm}^2\textrm{ at}\;\;T=0.93\, T_c \;\; ({\bf p}_{\mathrm{h}})\nonumber
 \end{align}
\section{Conclusions}

We have given the results of an improved calculation of the free pion form
factor ($T=0$) for low masses. It was found that it could be parametrised quite
well by the VMD-model. The
calculated form factor is seen to come close to the experimental data,
leaving only a small gap. We thus showed that it is possible to
calculate a highly non-perturbative quantity like a form factor, to
high qualitative agreement with experiment. The radius was found to be
close to the measured value, with a deviation of only $10\%$.

Whether a further lowering of the quark mass alone will resolve the
last gap with experiment, is uncertain. It might be necessary to
utilize chirally improved actions for a clarification of this point.
In any case, the results need to be approved by analyses including
dynamical quarks.

Rather, one should also
improve the methods, \textit{e.g.} by including dynamical quarks
(unquenched), or by using chiral symmetric actions.

Furthermore, we have presented the first finite temperature
calculation of the form factor using lattice QCD. It was found that up
to temperatures close to $T_c$, the situation is very similar to that
at $T=0$. The masses were found to be slightly lower at $0.93\, T_c$,
but the significance of the difference is small, although the effect
is systematic.

The structure of the pion-photon matrix element at finite temperature
may differ from that at $T=0$. Instead of one, we now can have two
independent form factors. Moreover, the form factors may depend on
more kinematic variables. We have investigated some of these
dependencies and found that the form factor decreases with decreasing
pion momenta. The differences are small and error bars overlap;
nonetheless, the effect is systematic.

The extracted radii confirm this picture; at finite temperature, they
are seen to depend on the pion momentum. In comparison with $T=0$, the
radius of the pion with the lower momentum is slightly larger.
Although error bars overlap, this increase in the radius might indicate
a change in the behaviour of the strong force, indicating the imminent
phase transition.

An obvious next step will be to calculate the form factor at
temperatures (just) above $T_c$, in order to search for evidence
whether deconfinement is complete, or if residual interactions, or
even real bound states still exist in that regime.

\section*{Acknowledgements}
The work of J.v.d.H and J.H.K. is part of the research program of the
Foundation for Fundamental Research of Matter (FOM) and the National
Organization for Scientific Research (NWO) of The Netherlands. The
research of E.L. is partly supported by Deutsche
Forschungsgemeinschaft (DFG) under grant FOR 339/2-1. The computations
were performed at the John von Neumann Institute for Computing (NIC),
J\"ulich and at SARA, Amsterdam under grant SG-119 by the Foundation
for National Computing Facilities (NCF).


\frenchspacing
\begin{thebibliography}{99}
\bibitem{Martinelli:1988bh} G.~Martinelli and C.~T. Sachrajda {\em Nucl. Phys.} {\bf B306} (1988) 865.

\bibitem{Draper:1989bp} T.~Draper, R.~M. Woloshyn, W.~Wilcox, and K.-F. Liu {\em Nucl. Phys.} {\bf
  B318} (1989) 319.

\bibitem{Luscher:1997ug} M.~L{\"u}scher, S.~Sint, R.~Sommer, P.~Weisz, and U.~Wolff {\em Nucl. Phys.} {\bf B491} (1997) 323--343,
  {{\tt hep-lat/9609035}}.

\bibitem{Sheikholeslami:1985ij}
B.~Sheikholeslami and R.~Wohlert {\em Nucl. Phys.} {\bf B259} (1985) 572.

\bibitem{Luscher:1997jn}
M.~L{\"u}scher, S.~Sint, R.~Sommer, and H.~Wittig {\em Nucl. Phys.} {\bf B491}
  (1997) 344--364, {{\tt hep-lat/9611015}}.

\bibitem{vanderHeide:2003kh}
J.~van~der Heide, J.~H. Koch, and E.~Laermann {\em Phys. Rev.} {\bf D69} (2004)
  094511, {{\tt hep-lat/0312023}}.

\bibitem{Bowler:1999ae}
{\bf UKQCD} Collaboration, K.~C. Bowler {\em et.~al.} {\em Phys. Rev.} {\bf
  D62} (2000) 054506, {{\tt hep-lat/9910022}}.

\bibitem{Bebek:1978pe}
C.~J. Bebek {\em et.~al.} {\em Phys. Rev.} {\bf D17} (1978) 1693.

\bibitem{Volmer:2000ek}
{\bf The Jefferson Lab F(pi)} Collaboration, J.~Volmer {\em et.~al.} {\em Phys.
  Rev. Lett.} {\bf 86} (2001) 1713--1716,
  {{\tt nucl-ex/0010009}}.

\end{thebibliography}
\end{document}